\documentclass[11pt]{article}
\usepackage[margin=1in]{geometry}
\usepackage{graphicx}
\usepackage{booktabs}
\usepackage{hyperref}
\usepackage{amsmath}
\usepackage{amssymb}
\usepackage{array}
\usepackage{longtable}
\usepackage{tikz}
\usetikzlibrary{positioning,calc,arrows.meta,patterns,decorations.pathreplacing}

\title{Legible Consensus: Topology-Aware Quorum Geometry\\for Asymmetric Networks}
\author{Tony Mason\thanks{\texttt{fsgeek@\{cs.ubc.ca,gatech.edu,wamason.com\}}}}
\date{March 2026}

\begin{document}
\maketitle

\begin{abstract}
Quorum design over asymmetric topologies conflates two independent concerns: inter-tier obligation (which tiers must participate for cross-tier safety) and intra-tier replication (how each tier survives local failures). Flat quorums treat all nodes as interchangeable; when consensus fails, the structure does not reveal whether a tier was unreachable or a tier lost too many replicas. We show that mapping a crumbling-wall quorum construction to a physically tiered network separates these concerns and makes the protocol's failure modes \emph{legible}: an operator can determine which tiers retain global consensus capability from the wall structure and connectivity state alone, without runtime probing. Using a 10-node Earth/LEO/Moon/Mars topology as a magnifying glass, we confirm that three of four tiers retain global liveness during Mars conjunction blackout; only the disconnected tier loses it. Consensus latency at each tier equals the speed-of-light round-trip to Earth: 183~ms (Earth), 131~ms (LEO), 5.1~s (Moon). The wall also imposes a leadership cost gradient on Multi-Paxos elections that symmetric grid quorums cannot express. A comparison between sparse and full-coverage topologies separates wall obligations from network reachability as independent liveness constraints. All results are design-level; quorum intersection is verified exhaustively in TLA+. Source, data, and reproduction commands: \url{https://github.com/fsgeek/eidolon}.
\end{abstract}

\section{Introduction}

Consensus protocols are usually studied over networks where all participants are roughly equidistant. In practice, many systems span topologies with structured asymmetry: datacenters at different distances, edge nodes behind slow links, or, in the extreme case that motivates this paper, tiers separated by minutes to hours of light-travel time.

The naive approach to consensus over such a topology is a flat quorum: majority (or any uniform threshold) over all nodes. This forces the commit path through the slowest tier on every operation. Flexible Paxos~\cite{howard2016} showed that safety requires only cross-intersection between Phase~1 and Phase~2 quorum families, not universal majority. This decouples the two phases and opens a design space: Phase~2 (the hot commit path) can be small and fast while Phase~1 (the rare election path) can be large and slow, as long as they intersect.

But how should that design space be explored? Most Flexible Paxos deployments choose quorum sizes, a number for each phase, without considering the \emph{shape} of those quorums. Peleg and Wool~\cite{peleg1995} showed that quorum geometry can be engineered around combinatorial structure rather than simple counts, with crumbling-wall constructions (a quorum system where nodes are arranged in rows of varying width; a quorum reads from successive rows, and the wall ``crumbles'' as rows lose members to failure) that achieve high availability through asymmetric quorum families. This paper composes these two ideas: Flexible Paxos's intersection requirement with Peleg and Wool's wall construction, mapped onto a physically tiered network where the rows of the wall correspond to latency tiers.

The deeper problem with flat quorums is not performance but \emph{conflation}. Quorum design over asymmetric topologies involves two independent concerns: \emph{inter-tier obligation} (which tiers must participate in each phase for cross-tier safety) and \emph{intra-tier replication} (how each tier internally survives node failures for local durability). A flat quorum treats every node as an interchangeable member of a single obligation pool, regardless of the structural role its tier plays. Nodes separated by minutes of light-travel time are placed in the same quorum as nodes across the room. When a quorum fails to form, the failure is opaque---was it a coverage problem (a tier is unreachable) or a replication problem (too many nodes crashed within a tier)? The structure doesn't say.

The wall separates these concerns. Inter-tier obligation is encoded in the wall geometry: one witness per tier on the downward path, with safety anchored at the bottom row. Intra-tier replication is each tier's own concern, invisible to the wall. This separation produces a protocol whose failure modes are \emph{legible}. We call a quorum construction legible if an operator can determine which tiers retain global consensus capability by inspecting the wall structure and current connectivity state, without runtime probing or failure detection---an $O(\text{tiers})$ check rather than enumeration over quorum subsets.

We demonstrate this using Eidolon, a discrete-event Paxos simulator, over a 10-node topology spanning Earth (5~nodes), LEO (1), Moon (1), and Mars (3)---the 5/1/1/3 topology. The interplanetary setting is not the point; it is the magnifying glass. At 22~light-minutes, the conflation error in flat quorum design is physically undeniable; the same error exists at terrestrial scales but is masked by generous timeouts.

This paper makes five contributions:
\begin{enumerate}
\item \textbf{The crumbling-wall construction mapped to physical topology.} Per-tier Phase~1 quorum families where each tier reads downward through the wall. Phase~2 is the Earth row. Intersection holds because every Phase~1 contains at least one Earth node. The construction is verified exhaustively in TLA+ over the full 10-node topology (Section~\ref{sec:construction}).
\item \textbf{Legible liveness degradation.} Under Mars conjunction blackout, the wall crumbles from the top: only Mars loses global Phase~1 capability. Earth, LEO, and Moon retain it. Consensus latency at each tier equals the speed-of-light round-trip to Earth: 183~ms (Earth), 131~ms (LEO), 5.1~s (Moon). Section~\ref{sec:results} includes a flat-vs-wall calibration on the same topology.
\item \textbf{Wall obligations versus network reachability.} A sparse network topology where LEO sees only 3~of~5 Earth ground stations breaks LEO's liveness despite the wall permitting it. The wall determines quorum \emph{obligations}; the network determines \emph{achievability}. These are independent constraints that compose to determine liveness (Section~\ref{sec:reachability}).
\item \textbf{Crash tolerance and the weakest-link migration.} Relaxing Phase~2 from all-of-Earth to $k$-of-Earth introduces crash tolerance, but the weakest link migrates from the global quorum to the Earth-local quorum. Under two Earth crashes, the topology-aware global construction maintains 98\% success while the flat local construction drops below 50\% (Section~\ref{sec:relaxed}).
\item \textbf{Leadership cost gradient.} The wall imposes a monotonic cost hierarchy on Multi-Paxos leader election: exhaustive TLA+ enumeration shows Earth has 4.6$\times$ more valid Phase~1 quorums than Mars, giving Earth leadership more valid quorum configurations and thus more survivable crash patterns. Symmetric grid quorums cannot express this gradient (Section~\ref{sec:leadership}).
\end{enumerate}
Each empirical claim is tied to an executable artifact and output file (Appendix~A). All results are design-level; deployment would introduce orbital dynamics, antenna scheduling, and variable link quality not modeled here.

\section{Background}
\label{sec:background}

\subsection{Flexible Paxos}
Classic Paxos~\cite{lamport1998} uses majority quorums for both phases, which are Phase~1 (prepare/promise) and Phase~2 (accept/accepted), guaranteeing intersection via the pigeonhole principle.

Flexible Paxos~\cite{howard2016} observed that the intersection requirement is \emph{cross-phase}, not within-phase: any Phase~1 quorum must intersect any Phase~2 quorum, but Phase~1 quorums need not intersect each other, nor Phase~2 quorums each other. Formally, safety holds if $q_1 + q_2 > |N|$, where $q_1$ and $q_2$ are the quorum sizes for Phase~1 and Phase~2 respectively. This is the uniform special case where all quorums have the same size; the crumbling-wall construction generalizes this with per-tier quorum families of different sizes, where the cross-intersection property holds by construction rather than by arithmetic majority. The key consequence is that Phase~2 (the hot path, executed on every commit) can be small if Phase~1 (the cold path, executed only during leader election) is correspondingly large.

\subsection{Crumbling-Wall Quorum Systems}
Peleg and Wool~\cite{peleg1995} introduced crumbling walls as quorum constructions that achieve high availability through asymmetric structure. Nodes are arranged in rows of potentially different widths. A quorum is formed by selecting elements from successive rows, reading from top to bottom, with each row's width determining how many elements must be chosen. The ``crumbling'' refers to rows that may lose elements to failure, narrowing the paths through the wall.

The key property: any two paths through the wall must share at least one node, guaranteeing the intersection that consensus requires.

\subsection{Composing the Two Ideas}
Grid and crumbling-wall constructions~\cite{cheung1992,peleg1995} show that quorum geometry can be structured around combinatorial properties rather than simple counts. Flexible Paxos shows that the two phases of consensus can use different quorum families. Our construction composes these: the rows of the crumbling wall correspond to physical tiers of the network, and each tier's Phase~1 quorum family reads downward through the wall from that tier's position. Phase~2 uses only the bottom row (the fastest tier). The intersection guarantee follows from the wall structure: every path through the wall reaches the bottom row, and Phase~2 \emph{is} the bottom row.

\section{Construction}
\label{sec:construction}

\subsection{System Model}
Let the acceptor universe be
\[
N = E \cup L \cup U \cup M,
\]
where the tiers are pairwise disjoint and represent Earth, LEO, Moon, and Mars respectively. In the evaluated topology,
\[
|E| = 5,\quad |L| = 1,\quad |U| = 1,\quad |M| = 3,\quad |N| = 10.
\]
We call this the 5/1/1/3 topology. Tiers are ordered by latency: Earth is the fastest (bottom of the wall), Mars is the slowest (top). The network is asynchronous with configurable per-link delay and jitter. Acceptors are crash-stop and non-Byzantine. Links may be removed during scheduled blackout windows.

The simulator uses three logical consensus scopes:
\begin{itemize}
\item \textbf{Earth-local}: Flexible Paxos over the five Earth nodes ($q_1 = 4$, $q_2 = 2$).
\item \textbf{Mars-local}: majority quorum over the three Mars nodes ($q = 2$).
\item \textbf{Global}: the crumbling-wall construction defined below, over all ten nodes.
\end{itemize}

\subsection{Per-Tier Phase~1 Families}

\begin{figure}[t]
\centering
\begin{tikzpicture}[
    node/.style={draw, minimum width=0.7cm, minimum height=0.7cm, font=\footnotesize},
    tierlab/.style={font=\footnotesize\bfseries, anchor=east},
    arrow/.style={-{Stealth[length=3pt]}, thick},
    phase2box/.style={draw, fill=black!8, minimum width=0.7cm, minimum height=0.7cm, font=\footnotesize},
]
\node[phase2box] (e1) at (0, 0) {$e_1$};
\node[phase2box] (e2) at (0.9, 0) {$e_2$};
\node[phase2box] (e3) at (1.8, 0) {$e_3$};
\node[phase2box] (e4) at (2.7, 0) {$e_4$};
\node[phase2box] (e5) at (3.6, 0) {$e_5$};
\node[tierlab] at (-0.6, 0) {$T_0$: Earth};

\node[node] (l1) at (1.8, 1.1) {$\ell_1$};
\node[tierlab] at (-0.6, 1.1) {$T_1$: LEO};

\node[node] (u1) at (1.8, 2.2) {$u_1$};
\node[tierlab] at (-0.6, 2.2) {$T_2$: Moon};

\node[node] (m1) at (0.9, 3.3) {$m_1$};
\node[node] (m2) at (1.8, 3.3) {$m_2$};
\node[node] (m3) at (2.7, 3.3) {$m_3$};
\node[tierlab] at (-0.6, 3.3) {$T_3$: Mars};

\draw[arrow, color=red!70!black, densely dashed] (4.1, 3.3) -- node[right, font=\tiny, pos=0.5] {} (4.1, -0.15);
\node[font=\tiny, color=red!70!black, anchor=west] at (4.1, 3.0) {$\mathcal{Q}_1^{(3)}$: all tiers};

\draw[arrow, color=blue!70!black, densely dashed] (4.5, 2.4) -- (4.5, -0.15);
\node[font=\tiny, color=blue!70!black, anchor=west] at (4.5, 2) {$\mathcal{Q}_1^{(2)}$};

\draw[arrow, color=green!50!black, densely dashed] (4.9, 1.5) -- (4.9, -0.15);
\node[font=\tiny, color=green!50!black, anchor=west] at (4.9, 1) {$\mathcal{Q}_1^{(1)}$};

\draw[arrow, color=black!60, densely dashed] (5.3, 0.3) -- (5.3, -0.15);
\node[font=\tiny, color=black!60, anchor=west] at (5.3, 0.2) {$\mathcal{Q}_1^{(0)}$};

\draw[decorate, decoration={brace, amplitude=4pt, mirror}] (-0.15, -0.5) -- (3.75, -0.5)
    node[midway, below=4pt, font=\footnotesize] {$\mathcal{Q}_2 = \{E\}$ (Phase~2: hot path)};
\end{tikzpicture}
\caption{The crumbling-wall construction over the 5/1/1/3 topology. Each tier's Phase~1 quorum reads downward through the wall (dashed arrows). Phase~2 uses only the Earth row (shaded). The wall narrows from bottom to top: wider rows offer more quorum choices.}
\label{fig:wall}
\end{figure}
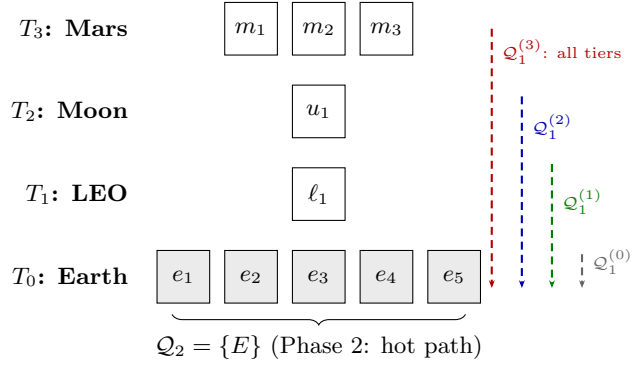

The wall maps tiers to rows, ordered bottom (Earth, fast) to top (Mars, slow), as shown in Figure~\ref{fig:wall}. We index tiers from the bottom: $T_0 = E$ (Earth), $T_1 = L$ (LEO), $T_2 = U$ (Moon), $T_3 = M$ (Mars). A proposer at tier~$i$ forms its Phase~1 quorum by reading \emph{downward} from row~$i$: it needs at least one respondent from its own tier and every tier below.

\begin{center}
\small
\begin{tabular}{llll}
\toprule
Initiating tier & Phase~1 scope & Tiers needed & Min size \\
\midrule
Earth (tier 0, bottom) & Earth only & 1 & 1 \\
LEO (tier 1) & LEO $+$ Earth & 2 & 2 \\
Moon (tier 2) & Moon $+$ LEO $+$ Earth & 3 & 3 \\
Mars (tier 3, top) & Mars $+$ Moon $+$ LEO $+$ Earth & 4 & 4 \\
\bottomrule
\end{tabular}
\end{center}

Let $\mathcal{Q}_1^{(i)}$ denote the Phase~1 family for a proposer at tier~$i$:
\[
\mathcal{Q}_1^{(i)} = \bigl\{Q \subseteq N \;\big|\; \forall\, j \le i:\; Q \cap T_j \neq \emptyset\bigr\}.
\]
A proposer at tier~$i$ needs at least one node from every tier at or below it ($j \le i$). For example, $\mathcal{Q}_1^{(2)}$ (Moon) requires intersection with $T_0$ (Earth), $T_1$ (LEO), and $T_2$ (Moon), which is every tier from Moon down to Earth. For relaxed Phase~2 (Section~\ref{sec:relaxed}), Phase~1 additionally requires enough Earth nodes to satisfy a pigeonhole intersection constraint.

\subsection{Phase~2 Family}

The Phase~2 family places the hot commit path entirely on the fast tier:
\[
\mathcal{Q}_2 = \{E\}.
\]
That is, Phase~2 requires all five Earth nodes (the strict construction). Section~\ref{sec:relaxed} relaxes this to $k$-of-$|E|$.

\subsection{Safety: Quorum Intersection}

\textbf{Proposition 1.} For every tier $i$ and every $Q_1 \in \mathcal{Q}_1^{(i)}$, $Q_2 \in \mathcal{Q}_2$: $Q_1 \cap Q_2 \neq \emptyset$.

\textit{Proof.} Every $\mathcal{Q}_1^{(i)}$ requires $Q_1 \cap T_0 \neq \emptyset$ (since $0 \le i$ for all $i$, i.e., Earth is at or below every tier). $Q_2 = E = T_0$. Therefore $Q_1 \cap Q_2 \supseteq Q_1 \cap T_0 \neq \emptyset$. \hfill $\square$

The intersection follows from the wall's structure: every path through the wall reaches the bottom row, and Phase~2 is the bottom row.

We verified this in TLA+. The \texttt{ExhaustiveIntersection} specification checks strict $Q_2$, relaxed $Q_2$ ($k = 4$ and $k = 3$), and crash-fault constructions across all four tiers' $Q_1$ families (11{,}789 states, complete). No intersection failure was found.

For the full Paxos protocol (not just quorum intersection), we verify safety over a reduced 6-node topology ($3E + 1L + 1U + 1M$) that preserves all tier-membership relationships. Cross-intersection requires $Q_1 \cap Q_2 \neq \emptyset$; since every $Q_1$ contains at least one node from $T_0$ (Earth) and $Q_2 \subseteq T_0$, intersection holds regardless of $|T_0|$. Adding nodes to a tier cannot break this property (monotonicity), so the reduction preserves all safety properties. Liveness properties (crash tolerance, progress under partial failure) are topology-size-dependent and are evaluated only at full scale. Over this reduced topology, single-decree Paxos agreement was model-checked exhaustively (67M~states, no violation). For the full 10-node topology, the state space exceeds what TLC can exhaust: we explored 6.5 billion states over 24 hours with the frontier still expanding, finding no violation. The reduced-topology complete result, plus the monotonicity argument, is the primary safety argument; the partial full-scale exploration provides additional confidence. Specifications are in \texttt{tla/} (see \autoref{app:traceability}).

\subsection{Liveness: Reading the Wall}
\label{sec:liveness-wall}

The wall determines liveness as follows. A proposer at tier~$i$ can complete Phase~1 if and only if it can reach at least one node in every tier~$j$ where $j \le i$ (from its own tier down to Earth). Phase~2 succeeds if the proposer can reach enough Earth nodes (all five under strict $\mathcal{Q}_2$).

During Mars conjunction blackout, cross-tier links between Mars and the inner system are removed. The connectivity state is:
\begin{itemize}
\item Mars $\leftrightarrow$ \{Moon, LEO, Earth\}: \textbf{partitioned}.
\item Moon $\leftrightarrow$ LEO $\leftrightarrow$ Earth: \textbf{connected}.
\end{itemize}

Reading the wall against this connectivity:
\begin{itemize}
\item \textbf{Earth} ($i{=}0$, needs $T_0$): \textbf{works}. Trivially satisfied.
\item \textbf{LEO} ($i{=}1$, needs $T_0, T_1$): \textbf{works}. Both tiers reachable.
\item \textbf{Moon} ($i{=}2$, needs $T_0, T_1, T_2$): \textbf{works}. All three tiers reachable.
\item \textbf{Mars} ($i{=}3$, needs $T_0, T_1, T_2, T_3$): \textbf{blocked}. Cannot reach inner tiers.
\end{itemize}

\begin{figure}[t]
\centering
\begin{tikzpicture}[
    node/.style={draw, minimum width=0.65cm, minimum height=0.65cm, font=\footnotesize},
    dead/.style={draw, minimum width=0.65cm, minimum height=0.65cm, font=\footnotesize, fill=black!20, text=black!40},
    ok/.style={draw, minimum width=0.65cm, minimum height=0.65cm, font=\footnotesize, fill=green!12},
    tierlab/.style={font=\footnotesize\bfseries, anchor=east},
    arrow/.style={-{Stealth[length=3pt]}, thick},
]
\node[ok] (e1) at (0, 0) {$e_1$};
\node[ok] (e2) at (0.85, 0) {$e_2$};
\node[ok] (e3) at (1.7, 0) {$e_3$};
\node[ok] (e4) at (2.55, 0) {$e_4$};
\node[ok] (e5) at (3.4, 0) {$e_5$};
\node[tierlab] at (-0.5, 0) {$T_0$};

\node[ok] (l1) at (1.7, 1.0) {$\ell_1$};
\node[tierlab] at (-0.5, 1.0) {$T_1$};

\node[ok] (u1) at (1.7, 2.0) {$u_1$};
\node[tierlab] at (-0.5, 2.0) {$T_2$};

\node[dead] (m1) at (0.85, 3.0) {$m_1$};
\node[dead] (m2) at (1.7, 3.0) {$m_2$};
\node[dead] (m3) at (2.55, 3.0) {$m_3$};
\node[tierlab, text=black!40] at (-0.5, 3.0) {$T_3$};

\draw[thick, red, densely dashed] (-2.0, 2.5) -- (3.5, 2.5);
\node[font=\tiny\bfseries, color=red, anchor=west] at (3.6, 2.5) {blackout};

\draw[arrow, color=green!50!black] (4.1, 0.5) --  (4.1, -0.3);
\node[font=\tiny, color=green!50!black] at (3.9, 0.1) {$\checkmark$};
\draw[arrow, color=green!50!black] (4.4, 1.3) -- (4.4, -0.3);
\node[font=\tiny, color=green!50!black] at (4.2, 1) {$\checkmark$};
\draw[arrow, color=green!50!black] (4.7, 2.1) -- (4.7, -0.3);
\node[font=\tiny, color=green!50!black] at (4.5, 1.7) {$\checkmark$};

\draw[arrow, color=red!70!black] (3.4, 3.4) -- (3.4, 2.7);
\node[font=\large, color=red] at (3.7, 3) {$\times$};
\end{tikzpicture}
\caption{Liveness under Mars conjunction blackout. The partition (dashed line) disconnects $T_3$. Earth, LEO, and Moon can still read downward to form Phase~1 quorums (green arrows). Mars cannot reach any inner tier (red $\times$). The wall crumbles from the top.}
\label{fig:blackout}
\end{figure}
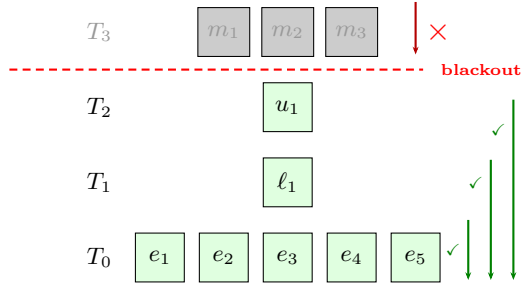

The liveness failure is scoped to exactly the disconnected tier. The wall crumbles from the top (Figure~\ref{fig:blackout}).

This is a \emph{legibility property}: given the wall structure and any connectivity state~$C$, the set of tiers retaining global Phase~1 capability is determined without runtime probing. For any tier~$i$, global Phase~1 succeeds if and only if every tier~$j \le i$ is reachable from~$i$ in~$C$, which is an $O(\text{tiers})$ check that requires no enumeration of quorum subsets. This is what distinguishes legibility from mere computability.

\subsection{What ``Global'' Means}

When an Earth-based proposer completes both phases using only Earth nodes, the decision is \emph{global} in the Paxos sense: any future proposer at any tier, upon completing Phase~1, will discover this value in the Earth acceptors' state and cannot contradict it. Global consensus does not require that all tiers participated, only that no future quorum can produce a conflicting result.

\section{Related Work}

Paxos established the standard safety argument for replicated state machines under asynchronous messaging and crash failures~\cite{lamport1998}. Flexible Paxos showed that safety requires only Phase~1/Phase~2 cross-intersection, not universal majority~\cite{howard2016}. Our construction builds directly on that result; the novelty is not in the intersection theorem but in mapping the quorum shape to physical topology and characterizing the resulting liveness properties per tier.

Grid and crumbling-wall quorum constructions~\cite{cheung1992,peleg1995} demonstrate that quorum geometry can be engineered around combinatorial structure. Howard et al.\ note explicitly that Flexible Paxos can use such constructions; we take up that suggestion. Peleg and Wool analyzed crumbling-wall availability under random, independent node failures. Our contribution is not the wall itself but the mapping: rows correspond to physical latency tiers, and the resulting per-tier liveness properties under scheduled, topology-correlated disconnection, are the properties the abstract framework identifies as possible but does not explore. Li, Chan, and Lesani~\cite{li2023} formalize heterogeneous quorum systems in which each process may use a different quorum family, and give necessary and sufficient conditions for liveness; our per-tier Phase~1 families are a concrete, topology-aligned instantiation of that general idea. Hierarchical quorum structures have a longer history in database replication~\cite{daudjee2004} and coordination services; our wall differs in that the hierarchy reflects physical latency tiers rather than organizational or logical grouping.

Geo-distributed consensus provides the closest practical analog. WPaxos~\cite{ailijiang2019wpaxos} uses Flexible Paxos with grid quorum layouts and per-object multi-leader coordination to keep commits local; it is the nearest point of comparison. WPaxos's grid quorums are symmetric: $Q_1$ spans $Z - f_z$ zones and $Q_2$ spans $f_z + 1$ zones, with the same structure regardless of which node initiates. Our per-tier $\mathcal{Q}_1^{(i)}$ families are asymmetric as the initiating tier determines the quorum scope, which produces a leadership cost gradient that symmetric grids structurally cannot express (Section~\ref{sec:leadership}). EPaxos~\cite{moraru2013} removes the leader bottleneck for non-conflicting commands. Mencius~\cite{mao2008} distributes leader responsibility across WAN sites. MDCC~\cite{kraska2013} achieves single-round cross-datacenter commits. All four optimize for tens-to-hundreds of milliseconds WAN latency and assume all replicas are usually reachable. Our setting differs in two respects: the asymmetry spans three orders of magnitude (sub-second to 1342~s), and the target phenomenon is scheduled total disconnection rather than heterogeneous latency.

Delay-tolerant networking~\cite{cerf2007,fall2003} motivates the operating regime but addresses routing and custody transfer, not consensus. CRDTs have been studied in DTN-like settings~\cite{guidec2023}. SwarmDAG~\cite{tran2019swarmdag} uses extended virtual synchrony for partition-tolerant robot swarms; Tran et al.~\cite{tran2019milcom} measure consensus latency under partitioning. Protocol Labs' IPC~\cite{delarocha2022} uses hierarchical subnets for blockchain scaling. NASA's DSN study~\cite{connell2011} evaluates database replication for ground stations. Our work occupies a gap: Paxos safety under topology-shaped quorum geometry and scheduled disconnection, with per-tier liveness characterization.

\section{Experimental Design}
\label{sec:expdesign}

Eidolon is a discrete-event simulator built on SimPy. Acceptors process Prepare and Accept messages with configurable processing delay. Proposers fan out messages to all acceptors and collect responses with a per-phase timeout. The network model assigns entities to named locations with per-link latency, jitter, and optional loss. Blackout is modeled as deterministic link removal; the repeater model substitutes a degraded alternate path. Event ordering is by simulation timestamp; the simulator is deterministic given a seed.

\subsection{Topology}

Five Earth ground stations (NA-West, Europe, Asia, SA-East, Africa) with cross-continental latencies of 30--120~ms. One LEO satellite with 20--45~ms links to ground stations. One Moon node at 1.28~s one-way. Three Mars nodes at configurable one-way delay (186--1342~s, representing closest approach to opposition). Mars-local links are 5~ms.

We evaluate two network variants:
\begin{itemize}
\item \textbf{Sparse}: LEO has links to 3 of 5 Earth stations (NA-West, Europe, Asia). Mars has links to 2 of 5 (NA-West, Europe). This models a single LEO satellite with limited ground coverage and a Mars relay with two tracking stations.
\item \textbf{Full coverage}: LEO and Mars have links to all 5 Earth stations. This models a LEO constellation with global coverage and a deep-space network with worldwide antenna placement.
\end{itemize}

\subsection{Experiments}

Three experiment families probe different aspects of the construction.

\begin{table}[t]
\centering
\small
\caption{Experimental design.}
\label{tab:design}
\begin{tabular}{p{0.14\linewidth}p{0.40\linewidth}p{0.38\linewidth}}
\toprule
Experiment & Purpose & Key parameters \\
\midrule
Parameter sweep & Confirm structural invariance of Earth-initiated global consensus across Mars delay and blackout duration & Mars delay $\in \{186, 750, 1342\}$~s; blackout $\in \{300, 900, 1800\}$~s; 50 seeds per point \\
Per-tier sweep & Measure which tiers retain global liveness; compare sparse vs.\ full-coverage topologies & Initiating tier $\in \{$Mars, Moon, LEO, Earth$\}$; both topologies; 186~s Mars delay; 900~s blackout; 50 seeds \\
Crash tolerance & Characterize the tradeoff between Phase~2 relaxation ($k$-of-$|E|$) and crash tolerance & $k \in \{3, 4, 5\}$; 0--2 Earth crashes; 186~s; 900~s blackout; 50 seeds \\
\bottomrule
\end{tabular}
\end{table}

Fixed parameters across all experiments are listed in Table~\ref{tab:params}. All seeds are 40--89 (50 runs per point). Results are reported as means with 95\% confidence intervals.

\begin{table}[t]
\centering
\small
\caption{Simulator parameters.}
\label{tab:params}
\begin{tabular}{llp{0.38\linewidth}}
\toprule
Parameter & Value & Notes \\
\midrule
Earth cross-continental delay & 30--120~ms & per link pair \\
LEO--ground delay & 20--45~ms & per ground station \\
Moon one-way delay & 1.28~s & fixed \\
Mars one-way delay & 186--1342~s & swept \\
Mars-local delay & 5~ms & fixed \\
Processing delay & 1~ms & per acceptor \\
Jitter & $\pm$10\% of link delay & uniform \\
Per-phase timeout & 500~ms & accommodates Earth-local RTT with margin; blackout effects are dominated by light-travel time, not timeout choice \\
Max Paxos rounds & 1 & per attempt \\
Blackout start & 600~s & simulation time \\
Simulation end & 4000~s & --- \\
Reconciliation interval & 120~s & between global attempts \\
Loss model & none & deterministic link removal for blackout \\
\bottomrule
\end{tabular}
\end{table}

\section{Results}
\label{sec:results}

\subsection{Flat versus Crumbling Wall: Geometry Alone Changes the Outcome}

To calibrate the effect of quorum geometry, we compare two constructions over the identical topology, blackout model, and parameter sweep. The \emph{flat} construction uses a single Phase~1 family requiring all four tiers:
\[
\mathcal{Q}_1^{\text{flat}} = \bigl\{Q \subseteq N \;\big|\; \forall\, j \in \{0,1,2,3\}:\; Q \cap T_j \neq \emptyset\bigr\}.
\]
Any blackout that removes a tier makes this quorum unformable. The \emph{crumbling wall} construction uses the per-tier Phase~1 families defined in Section~\ref{sec:construction}, with an Earth-based global proposer. Both satisfy cross-intersection with $\mathcal{Q}_2 = \{E\}$ (since both require $Q_1 \cap T_0 \neq \emptyset$).

\begin{table}[t]
\centering
\small
\caption{Effect of quorum geometry: flat vs.\ crumbling-wall Phase~1 over the same topology and blackout model. Earth-initiated global proposer, hard blackout, 50 seeds per point. All 18 parameter points (3 Mars delays $\times$ 3 blackout durations) return identical rates.}
\label{tab:flat-vs-wall}
\begin{tabular}{lrrr}
\toprule
Construction & During blackout & Post blackout & Avg latency \\
\midrule
Flat Phase~1 (all tiers required) & 0.0\% $\pm$ 0.0\% & varies by setting & varies \\
Crumbling wall (Earth reads Earth only) & 100.0\% $\pm$ 0.0\% & 100.0\% $\pm$ 0.0\% & 0.183~s \\
\bottomrule
\end{tabular}
\end{table}

The flat construction yields 0\% during-blackout success at every sweep point because it requires all tiers, including disconnected Mars. The crumbling wall yields 100\% because the Earth-based proposer's Phase~1 reads only Earth. Zero variance across 50 seeds confirms that outcomes here are structural consequences of quorum geometry and link state, not scheduler artifacts. We therefore treat this as calibration, not discovery.

\subsection{Per-Tier Liveness: The Wall Crumbles from the Top}

Table~\ref{tab:pertier} reports the central result. Under full network coverage, each tier runs its own global proposer during Mars conjunction blackout (186~s one-way, 900~s blackout, hard blackout model, 50 seeds).

\begin{table}[t]
\centering
\small
\caption{Per-tier global consensus during Mars conjunction blackout (full-coverage topology, hard blackout, 186~s Mars delay, 900~s blackout, mean $\pm$ 95\% CI over 50 seeds).}
\label{tab:pertier}
\begin{tabular}{llrrr}
\toprule
Initiating tier & Wall position & During-blackout & Avg latency & Recovery lag \\
\midrule
Earth & bottom (tier 0) & 100.0\% $\pm$ 0.0\% & 0.183~s $\pm$ 0.000 & 62.6~s $\pm$ 0.01 \\
LEO & tier 1 & 100.0\% $\pm$ 0.0\% & 0.131~s $\pm$ 0.000 & 61.8~s $\pm$ 0.01 \\
Moon & tier 2 & 100.0\% $\pm$ 0.0\% & 5.131~s $\pm$ 0.000 & 6.7~s $\pm$ 0.01 \\
Mars & top (tier 3) & 0.0\% $\pm$ 0.0\% & --- & --- \\
\bottomrule
\end{tabular}
\end{table}

Three of four tiers retain global consensus capability during hard Mars blackout. Only Mars, which is the disconnected tier, loses it. The liveness failure crumbles from the top of the wall, exactly as the construction predicts.

The latency column maps directly to physics. Moon's 5.131~s is two Paxos phases at the 1.28~s one-way Moon--Earth light delay ($\approx 4 \times 1.28$~s). Earth's 183~ms reflects cross-continental round-trips within the Earth tier. LEO's 131~ms is \emph{faster} than Earth, which is a counterintuitive result explained by the network topology: LEO-to-ground links (20--45~ms) are shorter than cross-continental Earth links (50--120~ms). Wall position determines quorum \emph{obligations}; the speed of light determines their \emph{cost}. These are correlated but not identical, and LEO is the proof.

Mars fails during blackout for the structural reason predicted by the wall (it cannot reach inner tiers for Phase~1). Mars also fails \emph{before} blackout, for a different reason: Mars-to-Earth round-trip time ($\approx$372~s per phase) exceeds the 500~s per-phase timeout budget for a complete two-phase Paxos round. This is not a blackout finding but a physics finding: Mars-initiated global consensus requires either patient timeouts or delegation to a lower-tier proposer.

Recovery lag differences between tiers (6.7~s for Moon vs.\ 62.6~s for Earth) are scheduling artifacts of the 120~s reconciliation interval interacting with blackout boundaries, not protocol properties.\footnote{Moon's slightly longer per-round time shifts its reconciliation cadence so the next attempt falls shortly after blackout ends. Earth's falls $\sim$62~s later. The effect is deterministic and independent of the consensus protocol.}

\subsection{Wall Obligations versus Network Reachability}
\label{sec:reachability}

Table~\ref{tab:sparse} repeats the per-tier experiment under the sparse network topology.

\begin{table}[t]
\centering
\small
\caption{Per-tier global consensus under sparse topology (same parameters as Table~\ref{tab:pertier}).}
\label{tab:sparse}
\begin{tabular}{llr}
\toprule
Initiating tier & Wall prediction & Sparse result \\
\midrule
Earth & works & 100.0\% $\pm$ 0.0\% \\
LEO & works & 0.0\% $\pm$ 0.0\% \\
Moon & works & 100.0\% $\pm$ 0.0\% \\
Mars & blocked & 0.0\% $\pm$ 0.0\% \\
\bottomrule
\end{tabular}
\end{table}

LEO drops from 100\% (full coverage) to 0\% (sparse). The wall says LEO needs LEO $+$ Earth; that obligation is satisfiable. But strict Phase~2 requires all five Earth nodes, and LEO has links to only three ground stations. The wall determines what a tier \emph{needs}; the network determines what it \emph{can reach}. Liveness requires both.

This makes legibility a two-step operator procedure: check the wall structure (which determines quorum obligations per tier) and check the network topology (which determines whether those obligations are achievable). The wall is the blueprint; the network is the building site.

Moon succeeds in both topologies because it has links to all five Earth ground stations. The sparse topology models a single LEO satellite with limited ground coverage; the three missing ground-station links are the difference between 0\% and 100\% for LEO-initiated consensus.

\section{Crash Tolerance and Coordinated Relaxation}
\label{sec:relaxed}

The strict Phase~2 choice $\mathcal{Q}_2 = \{E\}$ introduces a fragility: a single crashed Earth acceptor blocks all global Phase~2 progress. This section relaxes that choice.

\subsection{Relaxed Construction}
Let the relaxed Phase~2 family require $k$ of $|E|$ Earth nodes. We evaluate $k = 4$ (tolerates one crash) and $k = 3$ (tolerates two). Phase~1 requires correspondingly more Earth nodes to maintain intersection: a $k$-of-$|E|$ Phase~2 quorum omits at most $|E| - k$ Earth nodes, so any set of $|E| - k + 1$ Earth nodes intersects every Phase~2 quorum (pigeonhole). Phase~1 must therefore contain at least $|E| - k + 1$ Earth nodes. For $k = 4$ out of 5 Earth nodes, this means at least 2 Earth nodes in every $Q_1$; for $k = 3$, at least 3. Both constructions were verified exhaustively in TLA+.

The relaxed Phase~1 family for tier~$i$ is:
\[
\mathcal{Q}_{1,k}^{(i)} = \bigl\{Q \subseteq N \;\big|\; \forall\, j \le i:\; Q \cap T_j \neq \emptyset \;\;\text{and}\;\; |Q \cap E| \ge |E| - k + 1\bigr\}.
\]

\subsection{Results: The Weakest Link Migrates}
Table~\ref{tab:relaxed} reports the crash-tolerance sweep (Earth-initiated, repeater-assisted, 186~s Mars delay, 900~s blackout, 50 seeds).

\begin{table}[t]
\centering
\small
\caption{Coordinated relaxation under Earth crashes (mean $\pm$ 95\% CI, 50 seeds). Earth-local Flexible Paxos: ``std'' is $q_1{=}4, q_2{=}2$; ``maj'' is $q_1{=}q_2{=}3$.}
\label{tab:relaxed}
\begin{tabular}{llrrrrr}
\toprule
Global $\mathcal{Q}_2$ & Local & Crashes & Earth-local & Global during & Global post & Recovery (s) \\
\midrule
strict & std & 0 & 100.0\% & 100.0\% & 100.0\% & 488.9$\pm$2.6 \\
strict & std & 1 & 100.0\% & --- & 0.0\% & n/a \\
4-of-5 & std & 0 & 100.0\% & 100.0\% & 100.0\% & 489.9$\pm$2.5 \\
4-of-5 & std & 1 & 100.0\% & 100.0\% & 100.0\% & 489.1$\pm$2.1 \\
3-of-5 & std & 2 & 49.7\% & 98.0\%$\pm$3.9\% & 100.0\% & 512.7$\pm$2.0 \\
3-of-5 & maj & 2 & 100.0\% & 92.0\%$\pm$7.6\% & 100.0\% & 513.4$\pm$2.2 \\
\bottomrule
\end{tabular}
\end{table}

\textbf{Strict Phase~2 is completely crash-intolerant.} A single Earth crash produces permanent global liveness loss (row~2).

\textbf{The weakest link migrates.} At two crashes with standard Earth-local quorums, the \emph{global} construction ($k = 3$) maintains 98\% during-blackout success but the \emph{Earth-local} construction drops to 49.7\% because it cannot form a Phase~1 quorum of 4 with only 3 remaining Earth nodes (rows~5--6). The global construction's advantage is not that it spans more nodes---for the Earth-initiated proposer used here, both phases of the relaxed global construction touch only Earth nodes---but that its Earth requirements are smaller: at $k = 3$, both global phases need only the three surviving Earth nodes, while the Earth-local configuration still demands four. The structural insight is the legibility of the migration: which crashes matter depends on which quorum family's requirements they intersect, and the wall makes each family's requirements explicit.

\textbf{Coordinated relaxation restores resilience.} Relaxing the Earth-local quorum to majority ($q_1 = q_2 = 3$) recovers 100\% local success while maintaining 92\% global success (row~6). The tradeoff curve, with Phase~1 minima shown for the bottom-initiated (Earth) and top-initiated (Mars) proposers:

\begin{center}
\small
\begin{tabular}{lrrrl}
\toprule
Regime & Phase~1 min (E/M) & Phase~2 & Crash tol. & Commit cost \\
\midrule
Strict ($k = 5$) & 1 / 4 & 5-of-5 & 0 & 2-node Earth \\
Relaxed ($k = 4$) & 2 / 5 & 4-of-5 & 1 & 2-node Earth \\
Relaxed ($k = 3$) + maj & 3 / 6 & 3-of-5 & 2 & 3-node Earth \\
\bottomrule
\end{tabular}
\end{center}
An Earth-initiated $Q_1$ needs only the $|E| - k + 1$ Earth nodes; a Mars-initiated $Q_1$ adds one witness from each of the three upper tiers. Relaxation buys crash tolerance by growing the cold path, never the hot one: Phase~2 shrinks as Phase~1's Earth requirement grows.

\section{Threats to Validity and Limitations}

All results are design-level, not deployment-level.

\textbf{Anchor concentration.} The wall concentrates both safety and liveness at the bottom row. Every Phase~1 reads down to Earth; every Phase~2 is on Earth. In our construction, this concentration is sufficient for the legibility property: per-tier liveness determination is $O(\text{tiers})$ because a fixed reference point for cross-intersection eliminates the need to enumerate quorum subsets. Distributing the anchor across multiple rows would require enumerating which row combinations provide intersection, recovering the exponential checking that legibility eliminates. The single anchor is what makes the wall readable. Note, however, that the anchor is a \emph{tier}, not a node. Since safety depends only on tier membership (Section~\ref{sec:construction}), the anchor tier's internal replication is unconstrained by the wall---it can be scaled, geo-distributed, or backed by its own consensus protocol independently. The threat is tier-level partition (all of Earth unreachable simultaneously), not individual node failure; relaxed Phase~2 tolerates the latter.

\textbf{Abstract network model.} Blackout is deterministic link removal. Orbital dynamics, antenna scheduling, contention, signal coding, and variable link quality are not modeled. The LEO topology variant demonstrates that network reachability matters; a more realistic model with time-varying LEO coverage (orbital ground tracks) would show liveness fluctuating with satellite position.

\textbf{Stylized workload.} Reconciliation runs on a fixed 120~s interval. This creates scheduling artifacts in recovery lag measurements; recovery lag differences between tiers are workload properties, not protocol properties.

\textbf{Single topology.} We evaluate one topology size (5/1/1/3). The construction generalizes to other tier sizes, but we do not sweep Earth-tier size or Mars replication degree.

\textbf{Crash-stop only.} Byzantine behavior, correlated faults, reconfiguration during blackout, and storage failures are outside scope.

\textbf{Per-tier crash tolerance.} The interaction between wall position and crash tolerance (does a Moon-initiated proposer degrade differently from an Earth-initiated one under Earth crashes?) is not explored and remains future work.

\section{Discussion}
\label{sec:leadership}

\textbf{Separation of inter-tier obligation from intra-tier replication.} The separation introduced in Section~1 has a concrete cost consequence: the wall's per-tier Phase~1 obligation (one witness from the tier) is strictly cheaper than requiring intra-tier consensus. Mars's three nodes are independent witnesses---any one satisfies the wall's requirement. A Mars-internal majority would require two of three, adding a quorum obligation the wall does not need. The anchor tier is the sole exception: Phase~2 coherence requires it to behave as a single logical entity, achievable through internal replication (the standard 2PC coordinator trick). Non-anchor tiers benefit from redundancy without paying the cost of within-tier agreement. The TLA+ verification provides independent evidence for this separation: the reduced 6-node topology ($3E + 1L + 1U + 1M$) is not merely a computational shortcut but a model of the collapsed-tier design---shrinking Earth from five nodes to three preserves all quorum intersection properties precisely because safety depends on tier membership, not tier size. The same structural invariant that makes the reduction valid makes the deployment collapse cheap.

\textbf{Topology-scoped consistency.} During blackout, the wall separates tiers by which \emph{capabilities} they retain. Earth, LEO, and Moon can still form both global quorums: they can complete Phase~1 (discover every committed value---any Phase~1 quorum intersects every Phase~2 quorum, so nothing committed can be missed) and Phase~2 (extend the global log); under the standard Paxos client discipline, their operations on the global log remain linearizable. Mars retains neither capability: there are Earth-committed values Mars cannot discover until connectivity resumes, so it is confined to reading its local replica---a stale but locally ordered prefix of the global history. Mars is sequentially consistent locally but not linearizable with respect to the global log. The two capabilities are independent, and both matter: the Phase~1 condition (reaching every tier $j \le i$) lets a tier \emph{learn} the global history, while \emph{extending} it additionally requires reaching the Phase~2 quorum. Under the sparse topology of Section~\ref{sec:reachability}, LEO satisfies its Phase~1 obligation and can still learn the global history, but cannot reach the strict Phase~2 quorum and so cannot extend it. The wall tells you exactly which capabilities each tier holds, and hence which consistency regime it can sustain. At interplanetary distances, the gap between the global and local regimes is twenty-two minutes wide---the distinction that distributed systems students struggle with as an abstract definition becomes a physical fact you could set a timer by.

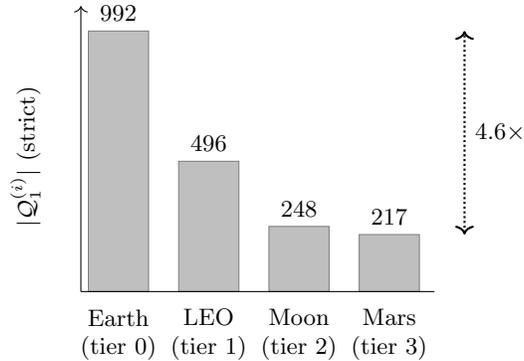
\begin{figure}[t]
\centering
\begin{tikzpicture}[
    bar/.style={fill=black!25, draw=black!50},
    barlab/.style={font=\footnotesize, anchor=south},
    ticklab/.style={font=\footnotesize},
]
\def\scale{0.0035}
\fill[bar] (0, 0) rectangle (0.8, 992*\scale);
\fill[bar] (1.2, 0) rectangle (2.0, 496*\scale);
\fill[bar] (2.4, 0) rectangle (3.2, 248*\scale);
\fill[bar] (3.6, 0) rectangle (4.4, 217*\scale);

\node[barlab] at (0.4, 992*\scale) {992};
\node[barlab] at (1.6, 496*\scale) {496};
\node[barlab] at (2.8, 248*\scale) {248};
\node[barlab] at (4.0, 217*\scale) {217};

\node[ticklab, anchor=north] at (0.4, -0.1) {Earth};
\node[ticklab, anchor=north] at (1.6, -0.1) {LEO};
\node[ticklab, anchor=north] at (2.8, -0.1) {Moon};
\node[ticklab, anchor=north] at (4.0, -0.1) {Mars};
\node[font=\footnotesize, anchor=north] at (0.4, -0.45) {(tier 0)};
\node[font=\footnotesize, anchor=north] at (1.6, -0.45) {(tier 1)};
\node[font=\footnotesize, anchor=north] at (2.8, -0.45) {(tier 2)};
\node[font=\footnotesize, anchor=north] at (4.0, -0.45) {(tier 3)};

\draw[->] (-0.1, 0) -- (-0.1, 3.8);
\node[font=\footnotesize, rotate=90, anchor=south] at (-0.5, 1.8) {$|\mathcal{Q}_1^{(i)}|$ (strict)};

\draw (-0.1, 0) -- (4.6, 0);

\draw[<->, thick, densely dotted] (5.0, 217*\scale) -- node[right, font=\footnotesize] {4.6$\times$} (5.0, 992*\scale);
\end{tikzpicture}
\caption{Phase~1 quorum count by initiating tier (strict $\mathcal{Q}_2$, from exhaustive TLA+ enumeration). Leadership flexibility increases monotonically from top to bottom of the wall.}
\label{fig:gradient}
\end{figure}

\textbf{Leadership hierarchy.} The wall imposes a cost gradient on Multi-Paxos leader election that symmetric quorum constructions cannot express. The exhaustive TLA+ enumeration quantifies this: Figure~\ref{fig:gradient} shows the number of valid Phase~1 quorums per tier. Earth has 4.6$\times$ more valid Phase~1 quorums than Mars. In Multi-Paxos, where leader election \emph{is} Phase~1, this means Earth leader election is not only faster (sub-second vs.\ minutes) but has more valid quorum configurations, increasing the number of crash patterns under which election can still succeed. The gradient is monotonic as leadership flexibility increases from top to bottom of the wall.

WPaxos~\cite{ailijiang2019wpaxos} uses symmetric grid quorums with uniform $Q_1$ structure regardless of initiator. By contrast, the wall uses tier-indexed families $\mathcal{Q}_1^{(i)}$, so partition failover cost is topology-proportional: if the leader is on Earth and Mars disconnects, leader election remains Earth-scoped.

\textbf{Terrestrial mapping.} The separation principle applies wherever topology creates structured latency asymmetry. Consider a 3-tier edge deployment: $T_0$ = cloud region (one logical anchor, internally replicated across availability zones), $T_1$ = metro edge (2~retail-store servers), $T_2$ = remote site (1~offshore-platform node). Phase~2 anchors at the cloud; each tier's Phase~1 reads downward. The cloud's internal replication is its own concern; the edge servers are independent witnesses, any one of which satisfies the wall's per-tier obligation. During planned WAN maintenance that disconnects the remote site, the wall predicts: cloud and metro edge retain global consensus; only the remote site loses it. Existing edge/cloud systems already separate these concerns in practice---cloud databases replicate within a region, edge nodes sync via ad hoc protocols---but they do so outside the quorum design, with bespoke consistency layers rather than structural guarantees. The wall gives that informal practice a formal foundation. The interplanetary setting makes the separation undeniable; terrestrial deployments would benefit from the same decomposition made explicit.

\textbf{Future work.} The separation principle suggests two natural extensions: scoped write leases (where the anchor tier delegates write authority to upper tiers over bounded slot ranges, with lease expiry on disconnection and conflict resolution via MRDTs/CRDTs), and empirical validation of the terrestrial mapping on edge/cloud topologies where the same decomposition applies but the latency asymmetry is less dramatic.

\section{Conclusion}

Flat quorums conflate inter-tier obligation with intra-tier replication. The crumbling wall separates them: one witness per tier on the downward path for cross-tier safety, while each tier's internal replication stays outside wall geometry. This yields legible liveness, a leadership cost gradient that symmetric constructions cannot express, and composable fault tolerance.

Over a 10-node interplanetary topology, we confirm that three of four tiers retain global consensus during hard Mars blackout, at latencies determined by the speed of light to Earth. The extreme asymmetry makes the conflation error in flat quorum design undeniable, but the separation principle applies wherever topology creates structured asymmetry---the wall gives a formal foundation to practices that edge/cloud systems already follow informally.

\appendix
\section{Claim-to-Artifact Traceability}
\label{app:traceability}
Table~\ref{tab:traceability} maps the paper's main claims to executable artifacts. Paths are relative to the repository root. Results were generated with Python~$\ge$3.14 and SimPy~$\ge$4.1.1. Persistent archive: Zenodo DOI \texttt{10.5281/zenodo.19122943}, commit \texttt{50fd0a3}.

\begin{longtable}{p{0.28\linewidth}p{0.30\linewidth}p{0.34\linewidth}}
\caption{Claim-to-artifact traceability.}
\label{tab:traceability} \\
\toprule
Claim & Evidence & Artifact(s) \\
\midrule
\endfirsthead
\toprule
Claim & Evidence & Artifact(s) \\
\midrule
\endhead
Flat construction: 0\% during blackout across all 18 sweep points. & Old construction sweep results. & \texttt{results/step9/} (prior sweep data). \\
Crumbling wall: 100\% during blackout across all 18 sweep points, 183~ms latency. & All rows show 100\% during-blackout. & \texttt{results/step9\_crumbling/\allowbreak step9\_sweep.csv}; \texttt{experiments/\allowbreak step9\_sweep.py}. \\
Per-tier liveness: 3 of 4 tiers retain global consensus during blackout. & Per-tier sweep, both topologies. & \texttt{results/tier\_liveness/\allowbreak tier\_sweep\_ci.csv}; \texttt{experiments/\allowbreak tier\_liveness\_sweep.py}. \\
Sparse topology: LEO drops to 0\%. & Sparse rows in per-tier sweep. & Same as above. \\
Weakest-link migration and coordinated relaxation. & Crash-tolerance sweep rows. & \texttt{results/step10/\allowbreak step10\_sweep\_ci.csv}; \texttt{experiments/\allowbreak step10\_sweep.py}. \\
TLA+ verification of quorum intersection. & Exhaustive enumeration, all tiers $\times$ all constructions. & \texttt{tla/\allowbreak ExhaustiveIntersection.tla}; \texttt{tla/\allowbreak QuorumIntersection.tla}. \\
Paxos agreement model-checked. & 67M states (reduced), 6.5B states (full, partial). & \texttt{tla/\allowbreak PaxosSmall.tla}; \texttt{tla/\allowbreak PaxosFull.tla}. \\
\bottomrule
\end{longtable}

\bibliographystyle{plain}
\bibliography{references}

@article{lamport1998,
  author = {Lamport, Leslie},
  title = {The Part-Time Parliament},
  journal = {ACM Transactions on Computer Systems},
  volume = {16},
  number = {2},
  pages = {133--169},
  year = {1998}
}

@inproceedings{howard2016,
  author = {Howard, Heidi and Malkhi, Dahlia and Spiegelman, Alexander},
  title = {Flexible Paxos: Quorum Intersection Revisited},
  booktitle = {20th International Conference on Principles of Distributed Systems (OPODIS 2016)},
  year = {2017}
}

@article{cheung1992,
  author = {Cheung, S. Y. and Ammar, M. H. and Ahamad, M.},
  title = {The Grid Protocol: A High Performance Scheme for Maintaining Replicated Data},
  journal = {IEEE Transactions on Knowledge and Data Engineering},
  volume = {4},
  number = {6},
  pages = {582--592},
  year = {1992}
}

@article{peleg1995,
  author = {Peleg, David and Wool, Avishai},
  title = {The Availability of Quorum Systems},
  journal = {Information and Computation},
  volume = {123},
  number = {2},
  pages = {210--223},
  year = {1995}
}

@techreport{cerf2007,
  author = {Cerf, Vinton and Burleigh, Scott and Hooke, Adrian and Torgerson, Leigh and Durst, Robert and Scott, Keith and Fall, Kevin and Weiss, Howard},
  title = {Delay-Tolerant Networking Architecture},
  institution = {RFC Editor},
  number = {RFC 4838},
  year = {2007}
}

@inproceedings{fall2003,
  author = {Fall, Kevin},
  title = {A Delay-Tolerant Network Architecture for Challenged Internets},
  booktitle = {Proceedings of the 2003 Conference on Applications, Technologies, Architectures, and Protocols for Computer Communications (SIGCOMM)},
  year = {2003}
}

@inproceedings{moraru2013,
  author = {Moraru, Iulian and Andersen, David G. and Kaminsky, Michael},
  title = {There Is More Consensus in Egalitarian Parliaments},
  booktitle = {Proceedings of the 24th ACM Symposium on Operating Systems Principles (SOSP)},
  pages = {358--372},
  year = {2013}
}

@inproceedings{mao2008,
  author = {Mao, Yanhua and Junqueira, Flavio P. and Marzullo, Keith},
  title = {{Mencius}: Building Efficient Replicated State Machines for {WANs}},
  booktitle = {Proceedings of the 8th USENIX Symposium on Operating Systems Design and Implementation (OSDI)},
  pages = {369--384},
  year = {2008}
}

@inproceedings{kraska2013,
  author = {Kraska, Tim and Pang, Gene and Franklin, Michael J. and Madden, Samuel and Fekete, Alan},
  title = {{MDCC}: Multi-Data Center Consistency},
  booktitle = {Proceedings of the 8th ACM European Conference on Computer Systems (EuroSys)},
  pages = {113--126},
  year = {2013}
}

@article{guidec2023,
  author = {Guidec, Fr\'{e}d\'{e}ric and Mah\'{e}o, Yves and No\^{u}s, Camille},
  title = {Supporting Conflict-Free Replicated Data Types in Opportunistic Networks},
  journal = {Peer-to-Peer Networking and Applications},
  volume = {16},
  pages = {395--419},
  year = {2023}
}

@article{tran2019swarmdag,
  author = {Tran, Jason A. and Ramachandran, Gowri S. and Shah, Parth M. and Danilov, Claudiu B. and Santiago, Roberto A. and Krishnamachari, Bhaskar},
  title = {{SwarmDAG}: A Partition Tolerant Distributed Ledger Protocol for Swarm Robotics},
  journal = {Ledger},
  volume = {4},
  number = {S1},
  year = {2019},
  pages = {25--31}
}

@inproceedings{tran2019milcom,
  author = {Tran, Jason A. and Ramachandran, Gowri S. and Danilov, Claudiu B. and Krishnamachari, Bhaskar},
  title = {An Evaluation of Consensus Latency in Partitioning Networks},
  booktitle = {MILCOM 2019 -- IEEE Military Communications Conference},
  year = {2019}
}

@inproceedings{delarocha2022,
  author = {de la Rocha, Alfonso and Kokoris-Kogias, Lefteris and Soares, Jorge M. and Vukoli\'{c}, Marko},
  title = {Hierarchical Consensus: A Horizontal Scaling Framework for Blockchains},
  booktitle = {IEEE 42nd International Conference on Distributed Computing Systems Workshops (ICDCSW)},
  year = {2022}
}

@techreport{connell2011,
  author = {Connell, Andrea M.},
  title = {An Analysis of Database Replication Technologies with Regard to Deep Space Network Application Requirements},
  institution = {Jet Propulsion Laboratory, California Institute of Technology},
  year = {2011}
}

@inproceedings{li2023,
  author = {Li, Yimeng and Chan, Andrew and Lesani, Mohsen},
  title = {Quorum Subsumption for Heterogeneous Quorum Systems},
  booktitle = {37th International Symposium on Distributed Computing (DISC 2023)},
  series = {LIPIcs},
  volume = {281},
  pages = {28:1--28:19},
  year = {2023}
}

@article{daudjee2004,
  author = {Daudjee, Khuzaima and Salem, Kenneth},
  title = {Lazy Database Replication with Ordering Guarantees},
  journal = {Proceedings of the 20th International Conference on Data Engineering (ICDE)},
  pages = {424--435},
  year = {2004}
}

@article{ailijiang2019wpaxos,
  title={Wpaxos: Wide area network flexible consensus},
  author={Ailijiang, Ailidani and Charapko, Aleksey and Demirbas, Murat and Kosar, Tevfik},
  journal={IEEE Transactions on Parallel and Distributed Systems},
  volume={31},
  number={1},
  pages={211--223},
  year={2019},
  publisher={IEEE}
}

\end{document}